\documentclass[12pt,a4paper]{article}
\pdfoutput=1
\usepackage[a4paper]{geometry}

\usepackage{amsmath}
\usepackage{amsfonts}
\usepackage{amssymb}
\usepackage{dsfont}
\usepackage[nosort]{cite}
\usepackage{hyperref}
\usepackage{multirow}

\newcommand{\dd}{\mathrm{d}}

\newcommand{\Op}{\mathcal{O}}

\newcommand{\zb}{\bar{z}}
\newcommand{\Lie}[2]{\mathcal{L}_{#1}#2}

\allowdisplaybreaks[2]
\numberwithin{equation}{section}	

\begin{document}

\vspace*{-1.5cm}
\begin{flushright}
  {\small
  LMU-ASC 03/21
  }
\end{flushright}

\vspace{1.75cm}

\begin{center}
{\LARGE
Holography and black holes\\ in asymptotically flat FLRW
}
\end{center}

\vspace{0.4cm}

\begin{center}
  Mart\'in Enr\'iquez Rojo, Till Heckelbacher
\end{center}

\vspace{0.3cm}

\begin{center} 
\textit{Arnold Sommerfeld Center for Theoretical Physics\\[1pt]
Ludwig-Maximilians-Universit\"at \\[1pt]
Theresienstra\ss e 37 \\[1pt]
80333 M\"unchen, Germany}\\[1pt]
\textit{Email:} \href{mailto:martin.enriquez@physik.lmu.de}{\texttt{martin.enriquez@physik.lmu.de}}, \\
\href{mailto:till.heckelbacher@physik.lmu.de}{\texttt{till.heckelbacher@physik.lmu.de}}
\end{center}

\vspace{1.8cm}


\begin{abstract}
\noindent
In this paper, we extend the treatment of asymptotically decelerating spatially flat FLRW spacetimes initiated in \cite{Rojo:2020zlz}. We show that a certain class of those metrics is ruled by the asymptotic algebra $\mathfrak{b}\mathfrak{m}\mathfrak{s}_s$, which belongs to a one-parameter family of deformations of $\mathfrak{b}\mathfrak{m}\mathfrak{s}$. Furthermore, we enlarge our ansatz to include $\text{Diff}(S^2)$ transformations whose asymptotic algebra $\mathfrak{g}\mathfrak{b}\mathfrak{m}\mathfrak{s}_s$ is a one parameter deformation of $\mathfrak{g}\mathfrak{b}\mathfrak{m}\mathfrak{s}$. Therefore, the holographic algebras $\mathfrak{b}\mathfrak{m}\mathfrak{s}_s$ and $\mathfrak{g}\mathfrak{b}\mathfrak{m}\mathfrak{s}_s$ in FLRW can be related to their flat counterparts through a cosmological holographic flow. Finally, we introduce a logarithmic ansatz in order to account for cosmological black holes, which does not generally satisfy the peeling property but preserves the asymptotic algebra.
\end{abstract}


\clearpage

\tableofcontents


\section{Introduction}

Asymptotic symmetries in decelerating spatially flat Friedmann– Lemaître– Robertson– Walker (FLRW) spacetimes at future null infinity $\mathcal{I}^+$ have been explored recently in \cite{Rojo:2020zlz,Bonga:2020fhx}. Both studies introduce a novel extension into a cosmological setting of the original work developed by Bondi, van der Burg, Metzner and Sachs in asymptotically flat spacetimes \cite{Bondi:1962px,Sachs:1962wk,Sachs:1962zza}. Rather surprisingly, the literature concerning the infrared structure of cosmological spacetimes is scarce. The first analysis of the asymptotic symmetry group of asymptotically FLRW spacetimes dates back to \cite{Hawking:1968qt}\footnote{This work suggests that the asymptotic symmetry group reduces to the global symmetry group of FLRW. Nevertheless, only pure gravitationally radiating, dust-filled universes with negative spatial curvature were considered, while our analysis deals with spatially flat universes allowing for general matter content. It would be interesting to apply modern techniques and compare the results with Hawking's analysis.}, although in the last few years some studies have been performed \cite{Mirbabayi:2016xvc,Shiromizu:1999iq,Bieri:2015jwa,Kehagias:2016zry,Tolish:2016ggo,Ferreira:2016hee,Chu:2016qxp,Chu:2016ngc,Bieri:2017vni,Hamada:2017gdg,Pajer:2017hmb,Donnay:2019zif,Chu:2021apx}. 

In particular, in \cite{Rojo:2020zlz} we obtained supertranslation and
superrotation-like asymptotic diffeomorphisms consistent with the
global symmetries of FLRW, computed how the asymptotic data is
transformed under them and studied the effect of these diffeomorphisms on pure FLRW, Sultana-Dyer, and cosmologically perturbed backgrounds. The present work aims to deepen into these findings by understanding the asymptotic symmetry algebra and extending our ansatz to include $\text{Diff}(S^2)$ transformations and cosmological black holes.

Asymptotically $\text{AdS}_3$ spacetimes were found to present a two dimensional conformal field theory (CFT) living on their boundary in \cite{Brown:1986nw}. That result was fundamental towards the formulation of the AdS/CFT correspondence \cite{Maldacena:1997re} and holography, the latter relating a quantum field theory (QFT) living in the boundary of a gravity theory to its gravitational bulk description. In this work, we show that the asymptotic algebra of decelerating spatially flat FLRW at $\mathcal{I}^{+}$ is given by $\mathfrak{b}\mathfrak{m}\mathfrak{s}_s\simeq(\mathfrak{w}\mathfrak{i}\mathfrak{t}\mathfrak{t}\oplus\mathfrak{w}\mathfrak{i}\mathfrak{t}\mathfrak{t})\ltimes_s\mathfrak{s}_s$ and related to its flat counterpart $\mathfrak{b}\mathfrak{m}\mathfrak{s}$ through a one-parameter family of deformations contained in $W(a,b,\bar{a},\bar{b})$ \cite{Safari:2019zmc,Safari:2020pje}, with $a=b=\bar{a}=\bar{b}=-\frac{1+s}{2}$. This constitutes a precise manifestation of a cosmological holographic flow from asymptotically flat ($s=0$) to asymptotically flat FLRW ($0<s<1$) spacetimes at $\mathcal{I}^{+}$ at the level of asymptotic algebras. 

Nevertheless, we are aware that in asymptotically flat spacetimes the inclusion of $\text{Diff}(S^2)$ transformations has been argued to be necessary in order to address the equivalence of the soft graviton theorems and the Ward identities for asymptotic symmetries \cite{Campiglia:2014yka,Campiglia:2015yka,Donnay:2020guq}. Therefore, we have allowed for general local $\text{Diff}(S^2)$ transformations and obtained their effect on the asymptotic data, as well as the extended asymptotic algebra $\mathfrak{g}\mathfrak{b}\mathfrak{m}\mathfrak{s}_s\simeq\mathfrak{v}\mathfrak{e}\mathfrak{c}\mathfrak{t}(S^2)\ltimes_{s}\mathfrak{s}_s$. It turns out the latter algebra is related to $\mathfrak{g}\mathfrak{b}\mathfrak{m}\mathfrak{s}$ in an analogous way that $\mathfrak{b}\mathfrak{m}\mathfrak{s}_s$ is to $\mathfrak{b}\mathfrak{m}\mathfrak{s}$, supporting the previous evidence for a cosmological holographic flow.

In addition, we realized that the Sultana-Dyer black hole was not covered by our ansatz in \cite{Rojo:2020zlz}. Here, we transform several inequivalent cosmological central inhomogeneities to Bondi coordinates and observe that we have to include logarithmic terms in $r$ to account for the black holes while the white hole solutions are included in the metrics described in \cite{Rojo:2020zlz,Bonga:2020fhx}. Interestingly enough, candidate metrics for primordial black holes \cite{2006PhDT........16M,Meissner:2021nvx} are now covered in our asymptotic metrics. This broader ansatz, which resembles the polyhomogeneous expansions in asymptotically flat black holes \cite{cite-key}, does not generally satisfy the $\frac{1}{r}$ peeling behaviour due to the presence of $\mathcal{O}(r^0)$ and $\mathcal{O}(\text{log}(r))$ terms in the Weyl tensor as $r\to\infty$. Nonetheless, it preserves the universal algebra $\mathfrak{g}\mathfrak{b}\mathfrak{m}\mathfrak{s}_s$ at infinity.

The structure of this paper is as follows: in section \ref{review}, we review the asymptotically decelerating spatially flat FLRW spacetimes studied in \cite{Rojo:2020zlz} and explore compatibility with the analysis performed in \cite{Bonga:2020fhx}. In section \ref{bmsk}, we identify the asymptotic symmetry algebra $\mathfrak{b}\mathfrak{m}\mathfrak{s}_s$ and show that it corresponds to a one-parameter family of deformations of $\mathfrak{b}\mathfrak{m}\mathfrak{s}$, revealing a cosmological holographic flow. In section \ref{diffS2_cosmo}, we enlarge our ansatz to include $\text{Diff}(S^2)$ diffeomorphisms, and in section \ref{gbmsk} we show that the asymptotic symmetry algebra $\mathfrak{g}\mathfrak{b}\mathfrak{m}\mathfrak{s}_s$ can be again understood as a one-parameter deformation of $\mathfrak{g}\mathfrak{b}\mathfrak{m}\mathfrak{s}_s$. In section \ref{bhmodels}, cosmological black holes are described in Bondi coordinates from the perspective of $\mathcal{I}^+$, and we include them in our ansatz by adding logarithmic terms in section \ref{polyhom}. Our conclusions are contained in section \ref{conclusion}, and we have collected the Lie derivatives for the asymptotic metrics in the appendix.

\section{Review of asymptotically spatially flat FLRW spacetimes}
\label{review}

In this section, we briefly review the treatment of asymptotically decelerating spatially flat FLRW universes at future null infinity $\mathcal{I}^+$ introduced in \cite{Rojo:2020zlz} and compare it with the analysis performed in \cite{Bonga:2020fhx}.

Spatially flat FLRW spacetimes are related to flat spacetimes by a Weyl transformation of the metric. In Bondi coordinates
\begin{equation}
	u=\eta-\sqrt{x^ix_i} \ , \ \ r=\sqrt{x^ix_i} \ , \ \ z=\frac{x^1+ix^2}{x^3+\sqrt{x^ix_i}} \ , \ \ \zb=\frac{x^1-ix^2}{x^3+\sqrt{x^ix_i}} \ ,
\end{equation}
this reads as
\begin{equation}
	\dd s^2=\left(\frac{r+u}{L}\right)^{2k}\left(-\dd u^2-2\dd u\dd r+\frac{4r^2}{(1+z\zb)^2}\dd z\dd\zb\right) \ ,
\end{equation}
where $\eta$ is the conformal time and $k=2/(3\omega+1)\geq0$, being $\omega=p/\rho$ the equation of state parameter of the fluid filling a decelerating universe.

To define spacetimes that asymptote to spatially flat FLRW we impose that they preserve the Bondi gauge  \cite{Bondi:1962px,Sachs:1962wk,Sachs:1962zza}
\begin{align}
    g_{rr}=0,&& g_{rA}=0,&& \partial_r\det\left(\frac{g_{AB}}{a^2 r^2}\right)=0 \ ,
\end{align}
where the indices $A,B\in\{z,\zb\}$ label the angular coordinates and $a=\left(\frac{r+u}{L}\right)^k$ is the scale factor. The determinant condition leaves some gauge freedom on the metric of the sphere. This condition can be strengthened, by fixing the determinant of the metric on the sphere entirely, which constrains the metric up to rescaling. Any changes of the spherical metric at leading order can by completely fixed by choosing the Bondi frame which restricts the angular metric to be the round metric on the sphere.

At the same time we allow for scalar, vector and tensor perturbations which do not spoil spatial homogeneity, isotropy and flatness and leave the matter content of the universe invariant in the limit $r\to\infty$. Another requirement is the closure of the expansion under asymptotic diffeomorphisms, meaning that such general transformations do not generate higher order terms in the $r$ expansion. Finally, we demand that the components and the trace of the Einstein tensor remain finite when dimensionally scaled.

In \cite{Rojo:2020zlz}, these considerations led us to study the asymptotic diffeomorphisms preserving the following ansatz
\begin{align}
	\dd s^2
	=&\left(\frac{r+u}{L}\right)^{2k}\left\{-\left(1-\Phi-\frac{2m}{r}\right)\dd u^2-2\left(1-\Psi-\frac{K}{r}\right)\dd u\dd r-2(r\Theta_A+U_A+\right. \nonumber \\
	&\frac{1}{r}N_A)\dd u\dd x^A+\left.\left((1+\Omega)r^2\gamma_{AB}+rC_{AB}+h_{AB}\right)\dd x^A\dd x^B\right\} \ .
	\label{eq:asymptpertspatflatFLRW}
\end{align}

It represents an expansion in $\frac{1}{r}$ for $r\to\infty$, where $\gamma_{z\zb}=\frac{2}{(1+z\zb)^2}$, $\gamma_{zz}=\gamma_{\zb\zb}=0$ is the round metric on the sphere and the expansion coefficients $\Phi,\Psi,\Omega,m,K,\Theta_A$, $U_A,N_A,C_{AB}$ and $h_{AB}$ are functions of $u,z$ and $\zb$. 

$\Phi,\Psi,\Omega,m$ and $K$ transform as scalars under spatial rotations while $\Theta_A, U_A$ and $N_A$ transform as vectors and $C_{AB}$ and $h_{AB}$ as tensors. There can be possible interrelations between some of those terms and some might actually turn out to produce physically unreasonable results. These are issues which can only be fixed by an on-shell treatment. By comparing the terms to the asymptotically flat expansions we expect the parameter $m$ to be related to the mass of a central inhomogeneity, $C_{AB}$ should be related to the gravitational radiation and $N_A$ to the angular momentum aspect of the spacetime. However, since our treatment is off-shell, a precise physical interpretation for each term can not be given so far. We pursue a first step into that direction in section \ref{bhmodels}, where we analyze some examples of cosmological black hole solutions.

In \cite{Bonga:2020fhx}, the authors proposed a different ansatz, using a more mathematical construction closer to the original BMS analysis \cite{Bondi:1962px,Sachs:1962zza,Sachs:1962wk}. The language they employed is based on the conformal completion \`a la Penrose \cite{Geroch:1977jn} and their assumptions were different to ours. They used a time-independent scale factor, which shifts the role of pure FLRW as the background metric but has the advantage to unveil a geometrical structure such that the spacetimes contained in their ansatz admit a cosmological null asymptote at infinity. That is indeed useful because it permits to identify (locally) a Bondi-like frame and a boundary behaving similarly to $\mathcal{I}^+$ in asymptotically flat spacetimes. This geometrical structure endows their spacetimes with an asymptotic symmetry algebra governing them, which they called $\mathfrak{b}_s\simeq\mathfrak{s}\mathfrak{o}(1,3)\ltimes\mathfrak{s}_s$, and it reduces to the original $\mathfrak{b}\mathfrak{m}\mathfrak{s}$ when $s=0$. It is a semi-direct sum of an algebra isomorphic to the Lorentz algebra (conformal global isometries of $S^2$) with an infinite-dimensional abelian algebra of supertranslation-like generators parametrized by conformally weighted functions on $S^2$. Nevertheless, the algebras $\mathfrak{b}_s$ for $s\neq0$ are not isomorphic to the BMS algebra.

Besides, they required the trace of the stress-energy tensor to be finite and its components to behave similarly to pure FLRW near to the boundary at null infinity. It turns out that the last requirement is too strong for our (more general) ansatz and one can easily check that the only compatible metrics with it are those satisfying $\Omega=\Psi=\delta\Omega=\delta\Psi=0$ \footnote{ \label{footy} $\Omega=\delta\Omega=0$ is a natural choice due to the remaining gauge freedom to fix the Bondi-frame \cite{Rojo:2020zlz}. However, $\Psi=\delta\Psi=0$ has a deeper meaning, related to the conformal structure at infinity \cite{Bonga:2020fhx,Geroch:1977jn,Wald:1984rg}.}. In this case, our asymptotic diffeomorphisms are given by \cite{Rojo:2020zlz}
\begin{align}
	\xi=&\xi^u(u,z,\zb)\partial_u+\left[r\xi^{r(V)}(z,\zb)+\xi^{r(0)}+\frac{1}{r}\xi^{r(1)}\right]\partial_r\nonumber\\ &+\left[V^{B}(z,\zb)+\frac1r\xi^{B(1)}+\frac{1}{r^2}\xi^{B(2)}\right]\partial_B \ ,
	\label{eq:Srotacc}
\end{align}
with
\begin{align}
    \xi^{r(V)}=-\frac{1}{2(1+k)}D_AV^A \ , \ \ \ 
    \xi^{u}=\frac{u}{2}\frac{(1+2k)}{(1+k)}D_AV^A+f(z,\zb) \ ,
    \label{eq:xiu}
\end{align}
\begin{align}
    \xi^{B(1)}=-D^B\xi^u \ , \ \ \  \xi^{B(2)}=\frac12\left(C^{AB}D_A\xi^u+KD^B\xi^u\right) \ ,
\end{align}
\begin{align}
    \xi^{r(0)}=&\frac{1}{1+k}\left[-\frac{1}{2}D_A\xi^{A(1)}-\frac{1}{2}\Theta^AD_A\xi^u+ku\xi^{r(V)}-k\xi^u\right] \ ,
     \\
    \xi^{r(1)}=&\frac{1}{2(1+k)}\left[C^A_B\Theta_AD^B\xi^u-2k\left(u^2\xi^{r(V)}-u\xi^{r(0)}-u\xi^u\right)\right. \nonumber\\
    &\left.-D_A\xi^{A(2)}-U^AD_A\xi^u\right] \ ,
\end{align}
and the asymptotic data transforms under them as \cite{Rojo:2020zlz}
\begin{align}
    \delta\Phi=&V^AD_A\Phi+\xi^u\partial_u\Phi-2\partial_u\xi^{r(0)}-2k(1-\Phi)\xi^{r(V)}
    -2(1-\Phi)\partial_u\xi^u\nonumber\\
    &+2\Theta_A\partial_u\xi^{A(1)}\label{eq:dNgenrot}\\
    \delta m=&\xi^u\partial_u m+V^AD_Am-k(1-\Phi)\xi^u-\left((1-2k)m-ku(1-\Phi)\right)\xi^{r(V)}\nonumber\\
    &-k(1-\Phi)\xi^{r(0)}+K\partial_u\xi^{r(0)}-\partial_u\xi^{r(1)}+m\partial_u\xi^u+U_A\partial_u\xi^{A(1)}\nonumber\\
    &+\frac12\xi^{A(1)}D_A\Phi+\Theta_A\partial_u\xi^{A(2)}\label{eq:dmgenrot}\\
    \delta K=&\xi^u\partial_u K+V^AD_AK+K\partial_u\xi^u-\Theta_A\xi^{A(1)}+2k\left(u\xi^{r(V)}-\xi^u-\xi^{r(0)}\right)\nonumber\\
    &+2kK\xi^{r(V)}\label{eq:dFgenrot}\\
    \delta C_{AB}=&\xi^u\partial_uC_{AB}+\Lie{V}{C_{AB}}+(1+2k)C_{AB}\xi^{r(V)}
    +\Lie{\xi^{C(1)}}{\gamma_{AB}}\nonumber\\
    &+\Theta_AD_B\xi^u+\Theta_BD_A\xi^u+2\gamma_{AB}((1+k)\xi^{r(0)}-ku\xi^{r(V)}+k\xi^u)\label{eq:dCABgenrot}\\
    \delta \Theta_A=&\xi^u\partial_u\Theta_A+\Lie{V}{\Theta_A}+(1+2k)\Theta_A\xi^{r(V)}
    -\partial_A\xi^{r(V)}+\Theta_A\partial_u\xi^u+\partial_u\xi^{(1)}_A\label{eq:dAAgenrot}\\
    \delta U_A=&\xi^u\partial_u\Theta_A+\Lie{V}{U_A}+\Lie{\xi^{C(1)}}{\Theta_A}+2k\Theta_A(\xi^u+\xi^{r(0)}-u\xi^{r(V)})\nonumber\\
    &-D_A\xi^{r(0)}+KD_A\xi^{r(V)}-(1-\Phi)D_A\xi^u+U_A\partial_u\xi^u+C_{AB}\partial_u\xi^{B(1)}\nonumber\\
    &+2kU_A\xi^{r(V)}+\Theta_A\xi^{r(0)}+\partial_u\xi_A^{(2)}\label{eq:dUAgenrot}\\
    \delta N_A=&\xi^u\partial_u N_A+\Lie{V}{N_A}-(1-2k)N_A\xi^{r(V)}+N_A\partial_u\xi^u
    +\Lie{\xi^{C(1)}}{U_A}+\Lie{\xi^{C(2)}}{\Theta_A}\nonumber\\
    &+KD_A\xi^{r(0)}-D_A\xi^{r(1)}+2mD_A\xi^u+2kU_A\left(\xi^{r(0)}+\xi^u-u\xi^{r(V)}\right)\nonumber\\
    &+2k\Theta_A\left(u^2\xi^{r(V)}-u(\xi^{r(0)}+\xi^u)+\xi^{r(1)}\right)+\Theta_A\xi^{r(1)}\nonumber\\
    &+h_{AB}\partial_u\xi^{B(1)}+C_{AB}\partial_u\xi^{B(2)}\label{eq:dNAgenrot} \ .
\end{align}

Indeed, we have checked that this restricted ansatz and diffeomorphisms are equivalent to those of \cite{Bonga:2020fhx} at the boundary $r\to\infty$. The main difference is that they use a non-dynamical scale factor (denoted by $\tilde{r}^{2k}$), while our scale factor $(r+u)^{2k}$ is dynamical. Nevertheless, if we factorize $r^{2k}$ in our scale factor and multiply the remaining part to the rest of the asymptotic metric expansion, both metrics coincide. The advantage of our approach is that the background metric is clearly pure FLRW, while their ansatz includes FLRW as an infinite expansion, but not as the base spacetime. This permits us to have a better intuition on the physical meaning of the different asymptotic data. Nevertheless, their approach turns out to be more suited to deal with transformations of metrics to Bondi coordinates and lets them identify the aforementioned geometrical structure \footnote{We will make use of both approaches in section \ref{polyhom}.}.

In order to make it easier for the reader who wants to compare both approaches, we display a rough dictionary between the coefficients in both works:
\begin{align}
    s=\frac{k}{1+k} \ , \ \
    F\leftrightarrow\xi^u \ , \ \ X^A\leftrightarrow V^A \ , \ \ U^A_{(1)}\leftrightarrow\Theta^A \ , \ \ \beta^{(1)}\leftrightarrow-\frac12 K \ . 
\end{align}
Using it, one can readily visualize a \textit{universal structure}. Our leading order algebra at $r\to\infty$, $r=\text{constant}$ coincides with that of \cite{Bonga:2020fhx} if one replaces $(1+s)\leftrightarrow\frac{1+2k}{1+k}$. In that limit, our diffeomorphisms become
\begin{align}
\xi[f(z,\zb),V^A(z,\zb)]=\left(f+\frac{u}{2}\frac{(1+2k)}{(1+k)}D_AV^A\right)\partial_u+V^A\partial_A \ ,
    \label{eq:universaldiffeo1}
\end{align}
leading to
\begin{align}
\xi[\hat{f},\hat{V}^A]=[\xi[f,V^A],\xi[f',V'^A]] \Rightarrow \nonumber \\ 
 \hat{f}=V^AD_Af'+\frac{(1+2k)}{2(1+k)}fD_AV'^A-V'^AD_Af-\frac{(1+2k)}{2(1+k)}f'D_AV^A \nonumber \\
 \hat{V}^A=V^BD_BV'^A-V'^BD_BV^A \ .
    \label{eq:algebra1}
\end{align}
Thus, we obtained the algebra $\mathfrak{b}\mathfrak{m}\mathfrak{s}_s\simeq(\mathfrak{w}\mathfrak{i}\mathfrak{t}\mathfrak{t}\oplus\mathfrak{w}\mathfrak{i}\mathfrak{t}\mathfrak{t})\ltimes_s\mathfrak{s}_s$ \footnote{We will explore $\mathfrak{b}\mathfrak{m}\mathfrak{s}_s$ in more detail in section \ref{bmsk}.}, which is equivalent to $\mathfrak{b}_s\simeq\mathfrak{s}\mathfrak{o}(1,3)\ltimes\mathfrak{s}_s$ of \cite{Bonga:2020fhx} when replacing $(1+s)\leftrightarrow\frac{1+2k}{1+k}$ and restricting to the six $V^{A}$ that are global conformal Killing vectors (CKV) on $S^2$.

As a result, the subset of metrics with $\Omega=\Psi=\delta\Omega=\delta\Psi=0$ is geometrically priviledged. Therefore, in order to unify and compatibilize treatments, we will restrict to it in what follows.

\section{$\mathfrak{b}\mathfrak{m}\mathfrak{s}_s$ as a deformation of $\mathfrak{b}\mathfrak{m}\mathfrak{s}$}
\label{bmsk}

Some aspects of the algebra $\mathfrak{b}\mathfrak{m}\mathfrak{s}_s$ have been explored in \cite{Bonga:2020fhx}. Here we take a different approach and expand it in terms of the basis of $z,\zb$ monomials on $S^2$. Our objectives are to express the algebra in terms of a more suited basis and to use it to relate it to the family of deformations of $\mathfrak{b}\mathfrak{m}\mathfrak{s}$, $W(a,b;\bar{a},\bar{b})$, discovered in \cite{Safari:2019zmc}. 

Taking into account that $\mathfrak{b}\mathfrak{m}\mathfrak{s}$ is proposed to govern flat holography, and $\mathfrak{b}\mathfrak{m}\mathfrak{s}_s$ appears to play a similar role in decelerating spatially flat FLRW holography, we wonder whether there exists a deformation relating both which could be interpreted as an $s$-cosmological holographic flow. Besides, the family of deformations $W(a,b,\bar{a},\bar{b})$ has been found to interpolate between $\mathfrak{b}\mathfrak{m}\mathfrak{s}$ ($W(-\frac{1}{2},-\frac{1}{2};-\frac{1}{2},-\frac{1}{2})$) and near-horizon symmetries ($W(a,a;a,a)$, \cite{Grumiller:2019fmp}) which are expected to play a major role in the description of black hole microstates. As a consequence, it is interesting to develop a similar analysis for $\mathfrak{b}\mathfrak{m}\mathfrak{s}_s$, which could eventually lead to the near horizon symmetry algebra for cosmological black holes.

Let us firstly define the basis of $z,\zb$ monomials on $S^2$:
\begin{equation}
    f_{mn}=\frac{z^m\zb^n}{1+z\zb} \ , \ \ V^z_m=-z^{m+1} \ , \ \ V^{\zb}_m=-\zb^{m+1} \ ,
\end{equation}
and the basis vectors $T_{mn}=\xi(f_{mn},0)$ $\mathcal{L}_m=\xi(0,V^z_m)$ and $\hat{\mathcal{L}_m}=\xi(0,V^{\zb}_m)$.
In terms of them, the non-vanishing commutators of (\ref{eq:algebra1}) become
\begin{align}
    [\mathcal{L}_m,\mathcal{L}_n]=(m-n)\mathcal{L}_{m+n} \ , \ \ [\hat{\mathcal{L}_m},\hat{\mathcal{L}_n}]=(m-n)\hat{\mathcal{L}}_{m+n} \ , \ \ [\mathcal{L}_m,\hat{\mathcal{L}_n}]=0 \ , \\
    [\mathcal{L}_m,T_{pq}]=\left[\frac{(m+1)}{2}(1+s)-p \right]T_{m+p,q}-s\frac{1}{1+z\zb}T_{m+p+1,q+1} \ ,  \\
    [\hat{\mathcal{L}_n},T_{pq}]=\left[\frac{(n+1)}{2}(1+s)-q \right]T_{p,q+n}-s\frac{1}{1+z\zb}T_{m+1,q+n+1} \ .
\end{align}
From the first commutators we obtain a $\mathfrak{w}\mathfrak{i}\mathfrak{t}\mathfrak{t}\oplus\mathfrak{w}\mathfrak{i}\mathfrak{t}\mathfrak{t}$ algebra. However, the last two commutators are more difficult to interpret. In fact, expanding $\frac{1}{1+z\zb}$, we observe that for $s\neq0$ the commutator does not finitely close, in the sense that we obtain infinitely many generators involved $T_{m+p+r,q+r}$ and $T_{p+r,q+n+r}$ with $r\in\mathbb{N}$. This already points to $s=0$ being a critical point of a flow. 

Nevertheless, if we use instead the basis of conformally weighted smooth functions on $S^2$ 
\begin{equation}
    \tilde{f}_{mn}=\frac{z^m\zb^n}{(1+z\zb)^{(1+s)}} \Rightarrow \ \tilde{T}_{pq}=\xi(\tilde{f}_{mn},0) \ ,
\end{equation}
we find
\begin{align}
    [\mathcal{L}_m,\tilde{T}_{pq}]=\left[\frac{(m+1)}{2}(1+s)-p \right]\tilde{T}_{m+p,q}  \\
    [\hat{\mathcal{L}_n},\tilde{T}_{pq}]=\left[\frac{(n+1)}{2}(1+s)-q \right]\tilde{T}_{p,q+n} \ ,
\end{align}
that is $\mathfrak{b}\mathfrak{m}\mathfrak{s}_s\simeq(\mathfrak{w}\mathfrak{i}\mathfrak{t}\mathfrak{t}\oplus\mathfrak{w}\mathfrak{i}\mathfrak{t}\mathfrak{t})\ltimes_s\mathfrak{s}_s$. The $\mathcal{L}_m$ act on $S^2$ as conformal Killing vectors and the operators $\tilde{T}_{pq}$ correspond to functions on $S^2$ with conformal weight $1+s$, that is an ideal of conformally weighted supertranslations which non-centrally extend the conformal algebra spanned by the $\mathcal{L}_m$.

It is clear that this algebra corresponds to a one-parameter deformation of $\mathfrak{b}\mathfrak{m}\mathfrak{s}$ in the generators $\tilde{T}_{pq}$ and the $[\mathcal{L}_m,\tilde{T}_{pq}]$, $[\hat{\mathcal{L}_n},\tilde{T}_{pq}]$ commutators. Taking into account that the parameter governing this deformation ($0\leq s=\frac{k}{1+k}<1$) is directly related to the equation of state of the base FLRW universe, that $\mathfrak{b}\mathfrak{m}\mathfrak{s}$ is expected to be the holographic algebra in flat holography and that $\mathfrak{b}\mathfrak{m}\mathfrak{s}_s$ plays the same role for asymptotically decelerating spatially flat FLRW \footnote{In fact, this is not completely right, as we will see later, a larger algebra $\mathfrak{g}\mathfrak{b}\mathfrak{m}\mathfrak{s}_s\simeq\mathfrak{v}\mathfrak{e}\mathfrak{c}\mathfrak{t}(S^2)\ltimes_{s}\mathfrak{s}_s$ is possibly the full holographic algebra, exactly as $\mathfrak{g}\mathfrak{b}\mathfrak{m}\mathfrak{s}$ is expected to be in flat holography \cite{Campiglia:2014yka,Campiglia:2015yka}. We will come back to this point in section \ref{diffS2_cosmo}.} ... this looks like a cosmological holographic flow deformation!

It turns out that the non-trivial deformations of $\mathfrak{b}\mathfrak{m}\mathfrak{s}$ have been studied in \cite{Safari:2019zmc,Safari:2020pje} and denoted by $W(a,b;\bar{a},\bar{b})$ with arbitrary $a,b\in\mathbb{R}$:
\begin{align}
    [\mathcal{L}_m,\mathcal{L}_n]=(m-n)\mathcal{L}_{m+n} \ , \ \ [\hat{\mathcal{L}_m},\hat{\mathcal{L}_n}]=(m-n)\hat{\mathcal{L}}_{m+n} \ , \ \ [\mathcal{L}_m,\hat{\mathcal{L}_n}]=0 \\
    [\mathcal{L}_m,\tilde{T}_{pq}]=-\left[p+bm+a \right]\tilde{T}_{m+p,q}  \\
    [\hat{\mathcal{L}_n},\tilde{T}_{pq}]=-\left[q+\bar{b}n+\bar{a} \right]\tilde{T}_{p,q+n} \ .
\end{align}
One can quickly realize that $\mathfrak{b}\mathfrak{m}\mathfrak{s}$ is given by $W(-\frac{1}{2},-\frac{1}{2};-\frac{1}{2},-\frac{1}{2})$ and $\mathfrak{b}\mathfrak{m}\mathfrak{s}_s$ is given by $W(-\frac{1+s}{2},-\frac{1+s}{2};-\frac{1+s}{2},-\frac{1+s}{2})$. Two concrete physically interesting cases correspond to radiation ($k=1\leftrightarrow s=\frac{1}{2}$, $W(-\frac{3}{4},-\frac{3}{4};-\frac{3}{4},-\frac{3}{4})$) and dust ($k=2\leftrightarrow s=\frac{2}{3}$, $W(-\frac{5}{6},-\frac{5}{6};-\frac{5}{6},-\frac{5}{6})$). For the decelerating range of $s$, these algebras are generic deformations of $\mathfrak{b}\mathfrak{m}\mathfrak{s}$ \cite{Safari:2020pje}, meaning that their deformations also lie in $W(a,b;\bar{a},\bar{b})$. 

As a final comment, note that due to symmetries shifting $a\leftrightarrow-a$ in $W(a,b;\bar{a},\bar{b})$, it might be indeed possible to relate $\mathfrak{b}\mathfrak{m}\mathfrak{s}_s$ to the algebra of accelerating spatially flat FLRW. Future studies of this algebra and their deformations shall be performed in order to unveil the exciting secrets of cosmological holography and its relation to flat holography \cite{Donnay:2020guq}, near horizon symmetries for cosmological black holes \cite{Donnay:2019zif}, fluid-gravity duality and membrane paradigm \cite{Penna:2015gza}, Virasoro extension $\hat{W}(-\frac{1+s}{2},-\frac{1+s}{2};-\frac{1+s}{2},-\frac{1+s}{2})$ and its deformations \cite{Safari:2019zmc,Safari:2020pje} ... and much more!

\section{$\mathrm{Diff}(S^2)$ transformations}
\label{diffS2_cosmo}

So far our ansatz only included conformal Killing vectors on the sphere. This was enforced by requiring that the leading order contribution to the metric on the sphere be given by the round metric.\\
If we want to allow for general diffeomorphisms on the 2-sphere, we have to relax that condition by allowing the diffeomorphisms to change the form of the metric on the sphere at leading order as well. The ansatz for the metric generalizes therefore to:
\begin{align}
	\dd s^2=&\left(\frac{r+u}{L}\right)^{2k}\left\{-\left(1-\Phi-\frac{2m}{r}\right)\dd u^2-2\left(1-\frac{K}{r}\right)\dd u\dd r-2(r\Theta_A+U_A+\right. \nonumber \\
	&\frac{1}{r}N_A)\dd u\dd x^A+\left.\left(r^2q_{AB}+rC_{AB}+h_{AB}\right)\dd x^A\dd x^B\right\} \ ,
	\label{eq:DiffS2metric}
\end{align}
where $q_{AB}$ is now a general metric on the sphere. Transforming this metric with the diffeomorphisms given in \eqref{eq:Srotacc}, we obtain the Lie derivatives given in appendix \ref{DiffS2Lie}. In this section, the indices of the quantities on the sphere are raised and lowered by $q_{AB}$ and $D_A$ denotes the covariant derivative with respect to $q_{AB}$.\\
To make the transformations consistent with Bondi gauge and the fall-offs required by our ansatz, we have to impose conditions on the coefficients of \eqref{eq:Srotacc}.\\
Bondi gauge requires that 
\begin{align}
	\Lie{\xi}{g_{rr}}=\Lie{\xi}{g_{rA}}=0&&
	\text{and}&&\partial_r\det\left(\frac{g_{AB}+\Lie{\xi}{g_{AB}}}{a^2r^2}\right)=0\label{eq:BondiGauge} \ .
\end{align}
The condition on $\Lie{\xi}{g_{rr}}$ is automatically fulfilled by the ansatz. The condition on $\Lie{\xi}{g_{rA}}$ gives the following restrictions:
\begin{align}
	\xi_A^{(1)}&=-D_A\xi^u \ , \\
	\xi_A^{(2)}&=\frac12\left(KD_A\xi^u-C_{AB}\xi^{B(1)}\right) \ .
\end{align}
To satisfy the determinant condition, we have to demand that $q^{AB}C_{AB}=0$ and that $q^{AB}S_{AB}=C^{AB}F_{AB}$, where $S_{AB}$ and $F_{AB}$ are defined in \eqref{eq:LieAB11srot}.\\
This leaves the leading order contribution to the spherical metric arbitrary, which means that the coefficient $\xi^{r(V)}$ in the expansion \eqref{eq:Srotacc} is a free parameter. A possible way to completely fix the gauge of our ansatz is to require that the determinant not only verifies \eqref{eq:BondiGauge} but that the determinant of the metric remains fixed under the diffeomorphisms. This leads to the requirement:
\begin{align}
	0=q^{AB}F_{AB}=4(1+k)\xi^{r(V)}+q^{AB}\xi^u\partial_u q_{AB}+2D_A V^A \ .
\end{align}
To solve this equation we demand that
\begin{align}
	\partial_u q_{AB}=0&&\text{and}&&\xi^{r(V)}=-\frac{1}{2(1+k)}D_AV^A \ .
\end{align} 
Physically this means that $q_{AB}$ contains no dynamical degrees of freedom.\\
We can now impose the fall-off conditions of our ansatz (\ref{eq:DiffS2metric}) to get further restrictions on the asymptotic transformations. The only additional requirements come from $\Lie{\xi}{g_{uA}}=\Op(r)$, $\Lie{\xi}{g_{uu}}=\Op(1)$ and $\Lie{\xi}{g_{ur}}=\Op(r^{-1})$. Together they give:
\begin{align}
	&\partial_uV^A=\partial_u\xi^{r(V)}=0\\
	&\partial_u\xi^u=-(1+2k)\xi^{r(V)}\qquad\Rightarrow\xi^u=f(z,\zb)+\frac{u}{2}\frac{1+2k}{(1+k)}D_AV^A \ .
\end{align}
We conclude that the asymptotic transformations and the asymptotic algebra are very similar to the superrotations given in \eqref{eq:universaldiffeo1}, with the only difference that $V^A$ are not required to be conformal Killing vectors but arbitrary $\mathrm{Diff}(S^2)$ transformations on the sphere.
This translates into the fact that the asymptotic algebra is now given by $\mathfrak{gbms}_s\simeq\mathfrak{vect}(S^2)\ltimes_{s}\mathfrak{s}_s$.

\section{$\mathfrak{g}\mathfrak{b}\mathfrak{m}\mathfrak{s}_s$ as a deformation of $\mathfrak{g}\mathfrak{b}\mathfrak{m}\mathfrak{s}$}
\label{gbmsk}

In section \ref{diffS2_cosmo}, we have extended our analysis to non-CKV on $S^2$, equivalent to (infinitesimal) $\text{Diff}(S^2)$ transformations, and we have found the asymptotic algebra $\mathfrak{g}\mathfrak{b}\mathfrak{m}\mathfrak{s}_s\simeq\mathfrak{v}\mathfrak{e}\mathfrak{c}\mathfrak{t}(S^2)\ltimes_{s}\mathfrak{s}_s$. Besides being expected to play a major role in holography for asymptotically spatially flat FLRW, this algebra constitutes, to our knowledge, the first deformation in the literature of $\mathfrak{g}\mathfrak{b}\mathfrak{m}\mathfrak{s}$. Both algebras constitute non-central extensions of $\mathfrak{v}\mathfrak{e}\mathfrak{c}\mathfrak{t}(S^2)$, the algebra of $\text{Diff}(S^2)$, which appears ubiquitously in several physical systems like fluids on the sphere \cite{10.5555/1965128}, membranes \cite{1982PhDT........32H,deWit:1988wri,Nicolai:1998ic}, flat holography \cite{Campiglia:2014yka,Campiglia:2015yka,Donnay:2020guq} and black hole entropy \cite{Donnay:2015abr}. As a consequence, it is of ultimate relevance to study this algebra.

Let us note the following before proceeding:

\begin{itemize}
    \item In this section we will show, following a similar analysis as in section \ref{bmsk}, how $\mathfrak{g}\mathfrak{b}\mathfrak{m}\mathfrak{s}_s$ looks in a better suited basis and how it relates exactly in the same way to $\mathfrak{g}\mathfrak{b}\mathfrak{m}\mathfrak{s}$ as $\mathfrak{b}\mathfrak{m}\mathfrak{s}_s$ to $\mathfrak{b}\mathfrak{m}\mathfrak{s}$, pointing to the existence of a similar family of deformations as the family $W(a,b;\bar{a},\bar{b})$ for $\mathfrak{b}\mathfrak{m}\mathfrak{s}$. Nevertheless, although no deformations of $\mathfrak{v}\mathfrak{e}\mathfrak{c}\mathfrak{t}(S^2)$ have yet been found \footnote{In a work in progress \cite{Enriquez-Rojo:2021rtv}, we are studying the possible deformations of $\mathfrak{v}\mathfrak{e}\mathfrak{c}\mathfrak{t}(S^2)$.}, it is well-known that (area-preserving part of) the algebra does not admit any central extension \cite{Bars:1988uj}. Therefore, it is very likely that there will not exist a family of deformations equivalent to $\hat{W}(a,b;\bar{a},\bar{b})$ for $\mathfrak{g}\mathfrak{b}\mathfrak{m}\mathfrak{s}$.
    
    \item We work in a local basis, which turns out to be over-complete for $\mathfrak{v}\mathfrak{e}\mathfrak{c}\mathfrak{t}(S^2)$ and singular at the poles. The situation is analogous to that of local superrotations where it was argued that the singularities could be understood in terms of cosmic string punctures \cite{Strominger:2016wns}. We are not aware of a similar interpretation for the basis we use in this section but it should be related because it still contains the de-Witt generators as a subalgebra.
    
    \item We use as base metric $\gamma_{AB}$, that is the round metric on $S^2$. This choice is non trivial in this case because the non-CKV on $S^2$ within local $\text{Diff}(S^2)$ transformations are allowed to change $\gamma_{AB}(z,\bar{z})\to q_{AB}(z,\bar{z})$.
\end{itemize}

Let us firstly define the basis of conformally weighted $z,\zb$ monomials on $S^2$:
\begin{equation}
    f_{mn}=\frac{z^m\zb^n}{(1+z\zb)^{(1+s)}} \ , \ \ V^z_{m,n}=-z^{m+1}\bar{z}^n \ , \ \ V^{\zb}_{m,n}=-z^{m}\zb^{n+1} \ ,
\end{equation}
and the basis vectors $T_{mn}=\xi(f_{mn},0)$ $\mathcal{L}_{m,n}=\xi(0,V^z_{m,n})$ and $\hat{\mathcal{L}_{m,n}}=\xi(0,V^{\zb}_{m,n})$.\\
In terms of them, the non-vanishing commutators of (\ref{eq:algebra1}) become
\begin{align}
    [\mathcal{L}_{m,n},\mathcal{L}_{r,s}]=(m-r)\mathcal{L}_{m+r,n+s} \ , \ \ [\hat{\mathcal{L}_{m,n}},\hat{\mathcal{L}_{r,s}}]=(n-s)\hat{\mathcal{L}}_{m+r,n+s} \ , \\ 
    [\mathcal{L}_{m,n},\hat{\mathcal{L}_{r,s}}]=-r\hat{\mathcal{L}}_{m+r,n+s}+n\mathcal{L}_{m+r,n+s} \ , \\
    [\mathcal{L}_{m,n},T_{pq}]=\left[\frac{(m+1)}{2}(1+s)-p \right]T_{p+m,q+n}  \ , \\
    [\hat{\mathcal{L}_{m,n}},T_{pq}]=\left[\frac{(n+1)}{2}(1+s)-q \right]T_{p+m,q+n} \ ,
\end{align}

that is $\mathfrak{g}\mathfrak{b}\mathfrak{m}\mathfrak{s}_s\simeq\mathfrak{v}\mathfrak{e}\mathfrak{c}\mathfrak{t}(S^2)\ltimes_s\mathfrak{s}_s$.

It is clear that this algebra corresponds to a one-parameter deformation of $\mathfrak{g}\mathfrak{b}\mathfrak{m}\mathfrak{s}$ in the $[\mathcal{L}_{m,n},\tilde{T}_{pq}]$, $[\hat{\mathcal{L}_{m,n}}, \tilde{T}_{pq}]$ commutators. The same comments at the end of section \ref{bmsk} apply here accordingly.

\section{Cosmological black holes}
\label{bhmodels}

Our aim in this section is to transform three inequivalent representatives of asymptotically spatially flat FLRW central inhomogeneities to Bondi coordinates. This will permit us to uncover a pattern for this class of solutions, which will motivate the logarithmic ansatz of section \ref{polyhom}.

Firstly, we consider the Thakurta solution \cite{Thakurta} which represents the late time attractor of a larger class of solutions, the so-called Faraoni-Jacques or generalized McVittie \cite{Faraoni:2007es,Hammad:2018hhv}. Besides, this metric was used in \cite{Meissner:2021nvx} to describe a potential model for primordial black holes. Next, we move on to Sultana-Dyer black and white holes \cite{Sultana:2005tp}, which have been studied in more detail in \cite{2006PhDT........16M} and also try to set the basis for describing primordial black holes which expand with the universe flow \cite{2006PhDT........16M}. Finally, we turn to Vaidya black and white holes \cite{Vaidya:1977zza}, representing inhomogeneities decoupled from the cosmological flow which aim to be a simplified model of astrophysical black holes. 

Some studies on the physical feasibility of these metrics have been performed \cite{2006PhDT........16M,Faraoni:2013aba,Hammad:2018hhv}, uncovering possible pathologies, like near horizon superluminality, or advocating doubts on whether or not they really represent black hole solutions. Nevertheless, they constitute the building blocks of potentially more realistic solutions (e.g. Lemaitre-Tolman-Bondi \cite{Firouzjaee:2008gs}) and should be included in our ansatz in the same way that the Schwarzschild solution belongs to asymptotically flat spacetimes.

\subsection{Thakurta black hole}

The non-rotating Thakurta black hole \cite{Thakurta} corresponds to superimposing a FLRW background over a Schwarzschild black hole in areal coordinates \footnote{This conformal time $\eta$ agrees with the one used in pure FLRW, being therefore the appropriate one to be compared with our asymptotic expansion.}:
\begin{align}
    \dd s^2&=-\left(1-\frac{2m}{r} \right)\dd t^2+a(t)^2\left[\frac{\dd r^2}{1-\frac{2m}{r}}+r^2\dd\Omega^2 \right] \nonumber \\
    &=a^2(\eta)\left[-\left(1-\frac{2m}{r} \right)\dd\eta^2+\frac{\dd r^2}{1-\frac{2m}{r}}+r^2\dd\Omega^2 \right] \ .
\end{align}
Using $\eta=u+r+2m\log\left(\frac{r}{2m}-1\right)$, we can write the previous metric in Bondi coordinates:
\begin{align}
    \dd s^2=&\left(\frac{u+r+2m\log\left(\frac{r}{2m}-1\right)}{L}\right)^{2k}\left[-\left(1-\frac{2m}{r}\right)\dd u^2-2\dd u\dd r+2r^2\gamma_{z\zb}\dd z\dd\zb\right] \ .
    \label{eq:ThakurthaBondi}
\end{align}

Before we continue, let us note that the metrics of the form
\begin{align}
    \dd s^2=\left(A\eta+B^2\eta^2\right)^2\left[-\left(1-\frac{2m}{r}\right)\dd\eta^2+\frac{4m}{r}\dd\eta\dd r+\left(1+\frac{2m}{r}\right)\dd r^2+2r^2\gamma_{z\zb}\dd z\dd\zb\right] \ 
\end{align}
have been proposed to describe primordial black holes in \cite{Meissner:2021nvx} and have the same transformation to Bondi coordinates in the large-$r$ regime, being clearly included in our logarithmic expansion of section \ref{polyhom} for $k=2$, after expanding the scale factor in series.

\subsection{Sultana-Dyer black hole}

The Sultana-Dyer solution \cite{Sultana:2005tp} consists of a time-dependent Kerr-Schild transformation of Minkowski. The resulting metric is given by:
\begin{equation}
    \dd s^2=a^2(\eta)\left[-\dd\eta^2+\dd r^2+r^2d\Omega+\frac{2m}{r}(\dd \eta\pm \dd r)^2 \right] \ ,
\end{equation}
where $\pm$ correspond respectively to black hole and white hole solutions \footnote{Note that this metric equals Vaidya when $ma^2(\eta)\to m$.}. The Sultana-Dyer black hole solution is equivalently written as:
\begin{align}
    \dd s^2=\left(\frac{\eta}{L}\right)^{2k}\left[-\left(1-\frac{2m}{r}\right)\dd\eta^2+\frac{4m}{r}\dd\eta\dd r+\left(1+\frac{2m}{r}\right)\dd r^2+2r^2\gamma_{z\zb}\dd z\dd\zb\right] \ .
\end{align}

In order to transform to conformal Schwarzschild, we have to reverse the Kerr-Schild transformation such that $d\bar{\eta}=d\eta-\frac{2m}{r-2m}dr$, $\bar{\eta}=\eta-2m\text{log}\left(\frac{r}{2m}-1 \right)$. Finally, to transform to Bondi coordinates, we use $\bar{\eta}\to u+r+2m\text{log}\left(\frac{r}{2m}-1 \right)$. As a result we obtain
\begin{align}
    \dd s^2=&\left(\frac{u+r+4m\log\left(\frac{r}{2m}-1\right)}{L}\right)^{2k}\left[-\left(1-\frac{2m}{r}\right)\dd u^2-2\dd u\dd r+2r^2\gamma_{z\zb}\dd z\dd\zb\right] \ .
    \label{eq:Sultana_Dyer1}
\end{align}

A similar analysis for the white hole reveals that the logarithms in the changes of coordinates cancel each other and we find:
\begin{align}
    \dd s^2=&\left(\frac{u+r}{L}\right)^{2k}\left[-\left(1-\frac{2m}{r}\right)\dd u^2-2\dd u\dd r+2r^2\gamma_{z\zb}\dd z\dd\zb\right] \ ,
    \label{eq:Sultana_Dyer2}
\end{align}
which is a solution in our expansion (\ref{eq:asymptpertspatflatFLRW}) \cite{Rojo:2020zlz}.

Let us finally comment that the metrics of this form, replacing the scale factor by a combination of radiation phase pasted to matter dominated phase, have been proposed to describe primordial black holes in \cite{2006PhDT........16M} and have an identical transformation to Bondi coordinates in the large-$r$ regime, being clearly included in our logarithmic expansion of section \ref{polyhom} for $k=2$, after expanding the scale factor in series.

\subsection{Vaidya black hole}

Vaidya's cosmological black and white holes \cite{Vaidya:1977zza} are obtained from application of a conformal transformation over Minkowski such that we obtain spatially flat FLRW in conformal coordinates, and then perform a time-independent Kerr-Schild transformation over it:
\begin{equation}
    ds^2=a^2(\eta)\left[-\dd\eta^2+\dd r^2+r^2\dd\Omega+\frac{2m}{ra^2(\eta)}(\dd \eta\pm \dd r)^2 \right] \ .
\end{equation}
The black hole solution is then written as:
\begin{align}
    \dd s^2=\left(\frac{\eta}{L}\right)^{2k}&\left[-\left(1-\frac{2m}{r\left(\frac{\eta}{L}\right)^{2k}}\right)\dd\eta^2+\frac{4m}{r\left(\frac{\eta}{L}\right)^{2k}}\dd\eta\dd r\right.\nonumber\\
    &\left.+\left(1+\frac{2m}{r\left(\frac{\eta}{L}\right)^{2k}}\right)\dd r^2+2r^2\gamma_{z\zb}\dd z\dd\zb\right] \ .
\end{align}

The exact transformation of this equation to Bondi coordinates is far from obvious to us. Nevertheless, in the limit $r\to\infty$, $\eta\sim r$ we obtain:
\begin{align}
    \dd s^2=a^2(r)\left[-\left(1-\frac{2m}{a^2(r)r}\right)\dd u^2-2\dd u\dd r+2r^2\gamma_{z\zb}\dd z\dd\zb\right] \ ,
\end{align}
upon solving the differential equation:
\begin{equation}
    \dd\eta=\dd r+\frac{4m}{\left(\frac{\eta(r)}{L}\right)^{2k}r-2m}\dd r \ ,
    \label{eq:unsolvable}
\end{equation}
in order to find $\eta(r)$ such that $a^{2}(r)=\left(\frac{\eta(r)}{L}\right)^{2k}$. Expanding around $r\to\infty$
\begin{equation}
    \dd\eta=\dd r+\frac{4m}{\left(\frac{\eta}{L}\right)^{2k}r}\left(1+\frac{2m}{\left(\frac{\eta}{L}\right)^{2k}r}+\frac{4m^2}{\left(\frac{\eta}{L}\right)^{4k}r^2}+... \right)\dd r \ ,
\end{equation}
we solve the first order expansion in the limit $\eta\sim r\to\infty$:
\begin{align}
    \dd\eta=\dd r+\frac{4m}{\left(\frac{\eta}{L}\right)^{2k}r}\dd r + ... \Rightarrow \  & \eta\sim r-\frac{2mL^{2k}}{k}r^{-2k} \ \ \  k\neq0 \nonumber \\ 
    &  \eta\sim r+4m\log(r) \ \ \  k=0 \ .
\end{align}
This permits us to check that Vaidya black hole is expressible in terms of logarithms and $k$-powers of $1/r$ in the region determined by $\eta\sim r\to\infty$, while it is not clear whether this metric is analytically expressible in terms of our ansatz in general \footnote{Note that one could try to solve (\ref{eq:unsolvable}) in the limit $\eta\sim r\to\infty$. Although, the solutions turn out to be very complicated hypergeometric functions, in the large $r$ regime one can observe logarithmic and polynomial behaviour in $\frac{1}{r}$.}.

An exact analysis for the white hole reveals that:
\begin{align}
    \dd s^2=&\left(\frac{u+r}{L}\right)^{2k}\left[-\left(1-\frac{2m}{\left(\frac{u+r}{L}\right)^{2k}r}\right)\dd u^2-2\dd u\dd r+2r^2\gamma_{z\zb}\dd z\dd\zb\right] \ ,
\end{align}
which turns out to be much simpler than the black hole but still presents subtleties because $k$ can be fractional and our $\frac{1}{r}$-expansion would not contain this example (exactly as in the black hole case). Nevertheless, the physically relevant cases of radiation and dust have $k=1$ and $k=2$ respectively, so both do not require any fractional expansion and are included in our ansatz (\ref{eq:asymptpertspatflatFLRW}) \cite{Rojo:2020zlz}.

Before moving on, we would like to comment on the possibility of primordial Vaidya black hole solutions. Metrics of this form, replacing the scale factor by a combination of radiation phase pasted to matter dominated phase, have been explored in \cite{2006PhDT........16M} and have a similar transformation to Bondi coordinates in the large-$r$ regime as the one explored in this section for $k=2$.

\vspace{1em}

\paragraph*{Physical interpretation}

These coordinate transformations highlight the fact that the ansatz for asymptotically spatially flat FLRW spacetimes presented in \cite{Bonga:2020fhx,Rojo:2020zlz} (section \ref{review}) covers only white hole solutions, whereas we observe more involved scale factors for the black holes which need a logarithmic ansatz (section \ref{polyhom}). This might be indeed related to the fact that white holes can be qualitatively regarded as an inversion of the arrow of time in black hole solutions, meaning that the black hole horizon is distinguished from $\mathcal{I}^+$ and its coupling to the cosmological flow manifests in the scale factor as a growing portion of spacetime, from which nothing can reach anymore $\mathcal{I}^+$. On the contrary, the white hole horizon has no effect on $\mathcal{I}^+$ more than its shared ``topological" $m$ contribution due to the singularity at $r=0$. 

In fact, we realize that the cases of Thakurta and Sultana-Dyer are similar, the only difference being the conformal time used to build them. In both the inhomogeneity expands with the universe, while Vaidya differs because it detaches from the expansion of the universe, possessing a shrinking event horizon and leading to complicated analytical dependence from the perspective of $\mathcal{I}^+$. 

It is also worth to note that Vaidya's metric points out the special role of $2k\in \mathbb{N}$ backgrounds which do not require a fractional $1/r$ expansion. Precisely the physically favoured radiation ($k=1$) and matter ($k=2$) dominated universes present this distinguished feature.

\section{Logarithmic expansion}
\label{polyhom}
As we saw in the analysis in section \ref{bhmodels} and in \cite{Rojo:2020zlz}, black hole models are not covered by the ansatz \eqref{eq:asymptpertspatflatFLRW}, since the expansion at $r\to\infty$ involves logarithmic terms. These logarithmic terms in the scale factor diverge towards $\mathcal{I}^+$. Since the retarded time $u$ is finite at null infinity, the log term will always dominate over the $u$ term in the scale factor and we, therefore, cannot write the metric at null infinity in the form of a time-dependent scale factor $a^2\propto(u+r)^{2k}$ times an asymptotically flat part.\\
Let us take as an example the Sultana-Dyer black hole \eqref{eq:Sultana_Dyer1} and factorize the scale factor in the following way:
\begin{align}
	a^2&=\left(\frac{u+r+4m\log\left(\frac{r}{2m}-1\right)}{L}\right)^{2k}
	=\left(\frac{r}{L}\right)^{2k}\left(1+\frac{u+4m\log\left(\frac{r}{2m}-1\right)}{r}\right)^{2k}\label{eq:log_scalefactor} \ .
\end{align}
The first term is divergent at $\mathcal{I}^+$, so we will extract it to be the asymptotic scale factor. The second part of \eqref{eq:log_scalefactor} is finite and differentiable at null infinity so we can expand it in terms of $\frac{\log^m r}{r^n}$:
\begin{align}
	\left(1+\frac{u+4m\log\left(\frac{r}{2m}-1\right)}{r}\right)^{2k}&=1+\frac{2k\left(u+4m\log\frac{r}{2m}\right)}{r}
	+\dots 
\end{align}
In order to generalize our ansatz to include cosmological black hole solutions, we conclude that we have to choose a time independent asymptote and expansion which includes logarithmic terms. This approach is similar to \cite{Bonga:2020fhx} apart from the logarithmic terms.\\
The most general ansatz we can write down, which includes logarithmic terms in the expansion and is finite and differentiable at null infinity, apart from an $r$-dependent scale factor, is given by:
\begin{align}
	\dd s^2=&\left(\frac{r}{L}\right)^{2k}\left\{-\left(1-\Phi-\frac{2m+A\log\frac{r}{B}}{r}\right)\dd u^2-2\left(1-\frac{K+E\log\frac{r}{F}}{r}\right)\dd u\dd r\right. \nonumber \\
	&+2\left(r\Theta_A+U_A+G_A\log\frac{r}{H_A}\right)\dd u\dd x^A\nonumber\\
	&\left.+\left(r^2q_{AB}+rC_{AB}+r M_{AB}\log\frac{r}{N_{AB}}\right)\dd x^A\dd x^B\right\}\label{eq:logansatz} \ ,
\end{align}
where $q_{AB}$ is a general metric on the sphere, such that the same arguments as for section \ref{diffS2_cosmo} apply. To preserve Bondi gauge, we now have to demand that:
\begin{align}
	q^{AB}C_{AB}=q^{AB}M_{AB}\log\frac{r}{N_{AB}}=0 \ .
\end{align}
To compare the asymptotic symmetry algebra of the above ansatz to the previous ansatz with a time dependent scale factor, we calculate the Lie derivatives of the above metric with respect to the diffeomorphisms from \eqref{eq:Srotacc} (see appendix \ref{logLie}).\\
If we impose the same gauge condition on the determinant of $q_{AB}$ as in section \ref{diffS2_cosmo} together with $\partial_u q_{AB}=0$, we discover that the asymptotic diffeomorphisms at leading order are exactly the same as the ones from section \ref{diffS2_cosmo}. The algebra is not spoiled by introducing logarithmic terms in the expansion and, therefore, applies to cosmological black holes as well.\\
As a final point we want to comment on the form of the Weyl tensor for metrics like \eqref{eq:logansatz}. For technical reasons, we consider as an example \eqref{eq:logansatz} with all the logarithmic terms given by $4m\log(r/2m)$ and the only other non-zero coefficient being $m(u,z,\zb)$. This corresponds to the asymptotic expansion of the Sultana-Dyer black hole with a time dependent mass.\\ For this case some non-vanishing components of the Weyl tensor are given by \footnote{We choose the same index convention as in \cite{Wald:1984rg}.}:
\begin{align}
	W_{ruru}&=\left(\frac{r}{L}\right)^{2k}\left(-\frac{2m}{r^3}-\frac{1}{r^4}\left(4km(u+4m\log(r/2m))\right)+...\right) \ , \\
	W_{ruAu}&=\left(\frac{r}{L}\right)^{2k}\left(\frac{3\partial_u m}{2r^2}+\frac{3kD_A m(u+4m\log(r/2m))}{r^3}+...\right) \ .
\end{align}
We can observe that the appearence of the logarithmic terms in the expansion spoils the peeling property of the Weyl tensor, building, therefore, a major difference with respect to the Schwarzschild solution within the asymptotically flat case.

\section{Summary and conclusions}
\label{conclusion}

In this paper, we delved into some properties arising from the novel asymptotically decelerating spatially flat FLRW spacetimes at $\mathcal{I}^+$ proposed in \cite{Rojo:2020zlz,Bonga:2020fhx}. Our major goals were to develop a better understanding of the geometry behind our ansatz (\ref{eq:asymptpertspatflatFLRW}) \cite{Rojo:2020zlz}, to obtain the asymptotic algebra ruling such asymptotic spacetimes, to extend our ansatz in order to admit cosmological black hole solutions and to allow not only for conformal Killing vectors (CKV) on $S^2$ but also for general local $\text{Diff}(S^2)$ transformations. On the way, we discovered a striking universal structure where the asymptotic algebra can be precisely related to the asymptotically flat one through a one-parametric family of deformations.

\subsection*{Summary of results}

Let us summarize our most important results:

\begin{itemize}
    \item We have compared the geometrical treatment and asymptotic metrics studied in \cite{Bonga:2020fhx} with (\ref{eq:asymptpertspatflatFLRW}). Even though they come from completely different approaches, the subset $\Omega=\Psi=\delta\Omega=\delta\Psi=0$ is equivalent at $r\to\infty$ to the metrics in \cite{Bondi:1962px}, while the incursion into the bulk appears to differ. Therefore, these metrics inherit a distinguished geometrical meaning and we restricted our analysis to them in this paper.
    
    \item The asymptotic algebra ruling the restricted set of metrics is given by $\mathfrak{b}\mathfrak{m}\mathfrak{s}_s\simeq(\mathfrak{w}\mathfrak{i}\mathfrak{t}\mathfrak{t}\oplus\mathfrak{w}\mathfrak{i}\mathfrak{t}\mathfrak{t})\ltimes_s\mathfrak{s}_s$, which reduces to $\mathfrak{b}_s\simeq\mathfrak{s}\mathfrak{o}(1,3)\ltimes\mathfrak{s}_s$ of \cite{Bonga:2020fhx}, when restricting the infinite set of superrotation generators to the six global CKV on $S^2$. 
    
    We deepened into its structure and found that $\mathfrak{b}\mathfrak{m}\mathfrak{s}_s$ is a one-parametric family of deformations, contained in the generic loci $a=b=\bar{a}=\bar{b}=-\frac{1+s}{2}$ of the broader family of deformations $W(a,b;\bar{a},\bar{b})$ of $\mathfrak{b}\mathfrak{m}\mathfrak{s}$ \cite{Safari:2019zmc,Safari:2020pje}. This result reveals a \textit{cosmological holographic flow} at the level of algebras, connecting flat ($s=0$) and decelerating spatially flat FLRW ($0<s<1$) spacetimes at $\mathcal{I}^+$.

    \item We augmented our ansatz to allow for local non-CKV on $S^2$, also called local $\text{Diff}(S^2)$ transformations, besides superrotations. We computed the asymptotic diffeomorphisms preserving these sets of asymptotic metrics and how they act on the asymptotic data. In addition, we investigated the asymptotic algebra governing these spacetimes at $\mathcal{I}^+$, $\mathfrak{g}\mathfrak{b}\mathfrak{m}\mathfrak{s}_s\simeq\mathfrak{v}\mathfrak{e}\mathfrak{c}\mathfrak{t}(S^2)\ltimes_{s}\mathfrak{s}_s$. The latter algebra is related to $\mathfrak{g}\mathfrak{b}\mathfrak{m}\mathfrak{s}$ in a similar way that $\mathfrak{b}\mathfrak{m}\mathfrak{s}_s$ is to $\mathfrak{b}\mathfrak{m}\mathfrak{s}$, supporting the previous evidence for a cosmological holographic flow and providing, to our knowledge, the first deformation of $\mathfrak{g}\mathfrak{b}\mathfrak{m}\mathfrak{s}$ in the literature. 
    
    \item We studied cosmological black and white hole solutions from the perspective of $\mathcal{I}^+$ using Bondi coordinates, noticing that white hole solutions are naturally included in the previous ansatz \cite{Rojo:2020zlz,Bonga:2020fhx} but black hole solutions require a logarithmic expansion. Moreover, we strikingly found in Vaidya's metric that only the physical cases $2k\in\mathbb{N}$ ($k=1$ radiation and $k=2$ dust) can be described using $\frac{1}{r^n}$-expansions with integer $n$. Otherwise, we would have to extend our ansatz to fractional $n$ expansions in order to include this central inhomogeneities detached from the flow. 
    
    \item From the study of these solutions, we built a logarithmic expansion in section \ref{polyhom} which includes the cosmological black holes and preserves the asymptotic algebra but presents many new unknown coefficients that are more difficult to interpret. Furthermore, this logarithmic ansatz does not satisfy in general the peeling property due to presence of $\mathcal{O}(r^0)$ (caused by the space filling fluid in FLRW, which is absent in flat spacetimes) and $\mathcal{O}(\text{log}(r))$ terms in the Weyl tensor. Although Thakurta and Sultana-Dyer solutions factorize the logarithm in the scale factor and verify the peeling property, after a generic asymptotic transformation, the new metrics do not satisfy it anymore.
\end{itemize}

\subsection*{Future research}

Finally, we briefly list some open questions and especially interesting research directions.

\begin{itemize}
    \item The treatment performed in this paper, as well as in previous works \cite{Rojo:2020zlz,Bonga:2020fhx}, is off-shell. This means that, although the gauge has been fixed, there is no direct and unique correspondence between the asymptotic coefficients and the, maximum six, degrees of freedom. An on-shell implementation, including asymptotic charges, for general relativity or alternative gravity theories is necessary towards practical quantitative applications and future testability. Unfortunately, it is not even clear how the global charges, mass, and angular momentum, can be globally defined for cosmological settings \cite{Firouzjaee:2010ia}, on the contrary to the ADM or Bondi charges in asymptotically flat spacetimes. 
    
    \item During the period of elaboration of this work, we tried to transform anisotropies which asymptote to spatially flat FLRW, like Thakurta-Kerr \cite{Thakurta,Mello:2016irl} or Bianchi Type I, to Bondi coordinates, in order to interpret them in our ansatz. We did not succeed, being the main reason that one must solve very difficult differential equations. Nonetheless, it would be illuminating to dispose of such metrics to analyze their properties. It would also be interesting to consider charged black holes, which would require an analysis of the asymptotic symmetries of the Maxwell field in FLRW background.
    
    \item Extension to accelerating spatially flat FLRW might be possible to achieve by means of indirectly using the properties of the deformation algebras $W(a,b;\bar{a},\bar{b})$ \cite{Safari:2019zmc,Safari:2020pje}. In fact, we speculate that the asymptotic symmetry algebra for accelerating spacetimes at the cosmological horizon might be given also by $\mathfrak{b}\mathfrak{m}\mathfrak{s}_s\simeq W(-\frac{1+s}{2},-\frac{1+s}{2};-\frac{1+s}{2},-\frac{1+s}{2})$ ($\mathfrak{g}\mathfrak{b}\mathfrak{m}\mathfrak{s}_s$ in the extended case) but more work is necessary to prove such a statement.
    
    \item The deformation relations found in \cite{Safari:2019zmc,Safari:2020pje}, between the near horizon algebras ($W(0,0;0,0)$ \cite{Donnay:2015abr}, $W(a,a;a,a)$ \cite{Grumiller:2019fmp}) and $\mathfrak{b}\mathfrak{m}\mathfrak{s}\simeq  W(-\frac{1}{2},-\frac{1}{2}; -\frac{1}{2},-\frac{1}{2})$,  make us hope that $\mathfrak{g}\mathfrak{b}\mathfrak{m}\mathfrak{s}_s$ plays a fundamental role for describing the microstates of cosmological black holes (and, possibly, cosmological horizons). Future research in this direction is highly encouraged.
    
    \item Prominent physical applications include the cosmological, especially the less studied scalar and vector, memory effects and scattering amplitudes in order to complete the cosmological infrared triangle \cite{Strominger:2014pwa,Strominger:2017zoo} and to dig into holography in cosmological spacetimes. 
\end{itemize}

As a final note, let us finish by pointing out two recent observations which might add phenomenological relevance to the metrics studied in this paper. Firstly, tension has been found against the standard cosmological paradigm asserting that we live in a FLRW with accelerated expansion (coming from the experimental results found in \cite{Riess:1998cb}). In \cite{Secrest:2020has,subir} it is suggested that the experimental data might actually be incompatible with the cosmological principle and, more concretely, with dark energy and accelerating expansion. This agrees with the work of \cite{Dvali:2020etd} which argues that the S-matrix formulation of quantum gravity excludes de-Sitter vacua. Secondly, candidate metrics for describing primordial black holes in \cite{2006PhDT........16M,Meissner:2021nvx} are included in our ansatz of section \ref{polyhom}. 

\vskip1em
\section*{Acknowledgements}

We would like to thank I.~Sachs for valuable comments regarding this work and general support, as well as Y.Z.~Chu, M.~Sheikh-Jabbari and K.~Prabhu for useful correspondence and to I.~Kharag for proofreading this paper. TH thanks the Hans-B\"ockler-Stiftung of the German Trade Union Confederation (DGB) for financial support. The work of MER was supported by the DFG Excellence Cluster ORIGINS.

\appendix
\numberwithin{equation}{section}

\section{Lie derivatives}

\subsection{$\text{Diff}(S^2)$ expansion}
\label{DiffS2Lie}
The Lie derivatives of \eqref{eq:DiffS2metric} with respect to the $\mathrm{Diff}(S^2)$ generators:
\begin{align}
a^{-2}\Lie{\xi}{g_{uu}}=&2r\left[\Theta^A\partial_u V_A-\partial_u\xi^{r(V)}\right] \nonumber\\
&+\left[V^AD_A\Phi+\xi^u\partial_u\Phi+2U_A\partial_uV^A-2\partial_u\xi^{r(0)}-2k(1-\Phi)\xi^{r(V)}\right. \nonumber\\
&\left.+2K\partial_u\xi^{r(V)}-2(1-\Phi)\partial_u\xi^u+2\Theta_A\partial_u\xi^{A(1)}\right] \nonumber \\
&+\frac{2}{r}\left[\xi^u\partial_u m-k(1-\Phi)\xi^u-\left((1-2k)m-ku(1-\Phi)\right)\xi^{r(V)}\right. \nonumber\\
&-k(1-\Phi)\xi^{r(0)}+V^AD_Am+\frac12\xi^{A(1)}D_A\Phi+K\partial_u\xi^{r(0)}-\partial_u\xi^{r(1)}\nonumber \\
&\left.+m\partial_u\xi^u+U_A\partial_u\xi^{A(1)}+\Theta_A\partial_u\xi^{A(2)}+N_A\partial_uV^A\right]+\Op(r^{-2}) \label{eq:uuDiff}\\
a^{-2}\Lie{\xi}{g_{ur}}=&\left[-(1+2k)\xi^{r(V)}-\partial_u\xi^u\right] \nonumber \\
&+\frac{1}{r}\left[\xi^u\partial_u K+V^AD_AK+K\partial_u\xi^u-\Theta_A\xi^{A(1)}\right. \nonumber \\
&\left.+2k\left(u\xi^{r(V)}-\xi^u-\xi^{r(0)}\right)+2kK\xi^{r(V)}\right]+\Op(r^{-2})\label{eq:urDiff} \\
a^{-2}\Lie{\xi}{g_{rA}}=&-q_{AB}\xi^{B(1)}-D_A\xi^u+\frac{1}{r}\left(KD_A\xi^u-C_{AB}\xi^{B(1)}-2q_{AB}\xi^{B(2)}\right)+\Op(r^{-2})\label{eq:rADiff}\\
a^{-2}\Lie{\xi}{g_{uA}}=&q_{AB}\partial_uV^B r^2+r\left[(1+2k)\Theta_A\xi^{r(V)}+\Lie{V}{\Theta_A}
-\partial_A\xi^{r(V)}+C_{AB}\partial_uV^B\right. \nonumber\\
&\left.+\xi^u\partial_u\Theta_A+\Theta_A\partial_u\xi^u+q_{AB}\partial_u\xi^{B(1)}\right] \nonumber \\
&+\left[(2k\Theta_A+\partial_u U_A)\xi^u+(1+2k)\Theta_A\xi^{r(0)}+2k\xi^{r(V)}(U_A-u\Theta_A)\right. \nonumber\\
&+\Lie{V}{U_A}+\Lie{\xi^{C(1)}}{\Theta_A}-D_A\xi^{r(0)}+KD_A\xi^{r(V)}-(1-\Phi)D_A\xi^u\nonumber \\
&\left.+h_{AB}\partial_uV^B+U_A\partial_u\xi^u+C_{AB}\partial_u\xi^{B(1)}+q_{AB}\partial_u\xi^{B(2)}\right]\nonumber \\
&+\frac{1}{r}\left[\xi^u\partial_u N_A+N_A\partial_u\xi^u+\Lie{V}{N_A}-(1-2k)N_A\xi^{r(V)}\right. \nonumber \\
&+KD_A\xi^{r(0)}-D_A\xi^{r(1)}+2mD_A\xi^u+2kU_A(\xi^{r(0)}+\xi^u-u\xi^{r(V)}) \nonumber \\
&+2k\Theta_A\left(u^2\xi^{r(V)}-u(\xi^{r(0)}+\xi^u)+\xi^{r(1)}\right)+\Theta_A\xi^{r(1)}+C_{AB}\partial_u\xi^{B(2)} \nonumber\\
&\left.+h_{AB}\partial_u\xi^{B(1)}+\Lie{\xi^{B(1)}}{U_A}+\Lie{\xi^{B(2)}}{\Theta_A}\right]+\Op(r^{-2}) \label{eq:uADiff}\\
a^{-2}\Lie{\xi}{g_{AB}}=&r^2F_{AB}+rS_{AB}+K_{AB} \label{eq:LieDiffAB2}
\end{align}
with 
\begin{align}
F_{AB}=&2(1+k)\xi^{r(V)}q_{AB}+\xi^u\partial_u q_{AB}+\Lie{V}{q_{AB}}\nonumber \ , \\
S_{AB}=&2q_{AB}((1+k)\xi^{r(0)}-ku\xi^{r(V)}+k\xi^u)+\Lie{\xi^{A(1)}}{q_{AB}}\nonumber\\
&+\Theta_AD_B\xi^u+\Theta_B D_A\xi^u+(1+2k)C_{AB}\xi^{r(V)}+\Lie{V}{C_{AB}}+\xi^u\partial_u C_{AB}\nonumber \ , \\
K_{AB}=&2kq_{AB}\left(u^2\xi^{r(V)}-u\xi^{r(0)}-u\xi^u\right)+2(1+k)q_{AB}\xi^{r(1)}+\Lie{\xi^{A(2)}}{q_{AB}}\nonumber\\
&+U_AD_B\xi^u+U_BD_A\xi^u+\Lie{\xi^{A(1)}}{C_{AB}}+2kh_{AB}\xi^{r(V)}+\xi^u\partial_u h_{AB}+\Lie{V}{h_{AB}}\label{eq:LieAB11srot} \ .
\end{align}

\subsection{Logarithmic expansion}
\label{logLie}
The Lie derivatives of \eqref{eq:logansatz} with respect to the $\mathrm{Diff}(S^2)$ generators:
\begin{align}
\Lie{\xi}{g_{uu}}=&\left(\frac{r}{L}\right)^{2k}\left\{\left(2\Theta_A\partial_uV^A-\partial_u\xi^{r(V)}\right)r+\Op(r^0)\right\}\\
\Lie{\xi}{g_{ur}}=&\left(\frac{r}{L}\right)^{2k}\left\{-(1+2k)\xi^{r(V)}-\partial_u\xi^u+\Op(r^{-1})\right\}\\
\Lie{\xi}{g_{rA}}=&\left(\frac{r}{L}\right)^{2k}\left\{-D_A\xi^u-q_{AB}\xi^{B(1)}+\Op(r^{-1})\right\}\\
\Lie{\xi}{g_{uA}}=&\left(\frac{r}{L}\right)^{2k}\left\{q_{AB}\partial_uV^B r^2+\Op(r)\right\}\\
\Lie{\xi}{g_{AB}}=&\left(\frac{r}{L}\right)^{2k}\left\{r^2\left(\Lie{V}{q_{AB}}+2(1+k)q_{AB}\xi^{r(V)}+\xi^u\partial_u q_{AB}\right)+\Op(r)\right\}
\end{align}

\clearpage
\nocite{*}
\bibliography{references}
\bibliographystyle{JHEP}


\end{document}